\documentclass[prd,aps,10pt,twocolumn,nofootinbib,superscriptaddress,showpacs]{revtex4-2}

\usepackage{amssymb,amsmath}
\usepackage{xcolor}
\usepackage[utf8]{inputenc}

\usepackage{enumitem}
\usepackage{empheq}
\usepackage{graphicx}
\usepackage{framed}
\usepackage[makeroom]{cancel}

\usepackage[makeroom]{cancel}
\usepackage[caption=false]{subfig}
\usepackage[colorlinks=true,
            citecolor=green,
            linkcolor=red,
            filecolor=cyan,
            urlcolor=magenta,
            backref=false]{hyperref}

\newcommand{\dd}{\mbox{d}}

\newcommand{\ts}[1]{{\boldsymbol{#1}}}

\begin{document}

\title{Carr criterion and mass gaps in non-singular primordial black hole formation}

\author{Jens Boos}
\email{jens.boos@kit.edu}
\affiliation{Institute for Theoretical Physics, Karlsruhe Institute of Technology, D-76128 Karlsruhe, Germany}

\author{Arif Ka\u{g}an G\"undo\u{g}du}
\email{gundogdu.arif@metu.edu.tr}
\affiliation{Department of Physics, Middle East Technical University, 06800 Ankara, T\"urkiye}

\author{Marek Hartenfels}
\email{marek.hartenfels@student.kit.edu}
\affiliation{Institute for Theoretical Physics, Karlsruhe Institute of Technology, D-76128 Karlsruhe, Germany}

\date{May 28, 2026}

\begin{abstract}

Non-singular gravitational theories are expected to be relevant in the early universe. In this paper, we derive a set of effective Friedmann equations describing the dynamics of matter shells in the presence of a gravitational regulator $\ell$. We find that such a regulator induces a primordial black hole mass gap such that below a certain mass $M_\text{gap}(\ell, R_H)$ no black holes can form. The order of magnitude of this mass gap is set by the regulator $\sim c^2\ell/G$, with subleading dependence on the horizon radius at time of formation $R_H$. Finally, we show that over a wide range of equation of state parameters $\omega = 0 \dots 1/3$, the mass gap implies a Carr criterion of the form $\delta_H > 2G M_\text{gap}/R_H - 1$. If the horizon size is of the same order of the regulator, $R_H \sim \ell$, this new criterion is stronger than the traditional Carr criterion for primordial black hole formation. This connects the primordial black hole abundance directly to the presence of gravitational regulators.

\end{abstract}

\maketitle

\section{Introduction}

As the direct observational evidence for the existence of black holes in the Universe grows, their role has changed from a prediction of general relativity to a promising avenue to search for traces of new physics \cite{Cardoso:2019rvt}. For example, the singularity problem in general relativity hints towards a fundamental ultraviolet incompleteness of the theory, which is also mirrored in its perturbative non-renormalizability. Adding terms to the Einstein--Hilbert action that are of higher order in curvature could potentially alleviate this \cite{Stelle:1976gc}, but the precise details of such theories remain under active investigation with no clear contender having as of yet emerged \cite{Salvio:2018crh,Bonanno:2020bil,Buoninfante:2024yth}.

Viewed from a different perspective, since it is likely that such higher-order curvature terms are present in any self-consistent singularity-free theory of gravity, a natural question to ask is for which black holes do such contributions play a role \cite{Lu:2015cqa}? Since the spacetime curvature $\mathcal{R}$ in the vicinity of a black hole of mass $M$ scales as $\mathcal{R} \sim c^4/(GM)^2$, small-mass black holes feature the largest spacetime curvature at their horizon. For this reason they are the most susceptible to possible deviations from general relativity.

This is mirrored in current experimental efforts relating to the direct imaging of supermassive black holes, which, while experimentally impressive, are not yet able to resolve possible deviations from general relativity \cite{Vagnozzi:2019apd,Khodadi:2020jij,EventHorizonTelescope:2021dqv}. On the other hand, primordial black holes are hypothetical objects \cite{Hawking:1971ei,Carr:1974nx,Carr:1975qj} that have come under increasing scrutiny in the last decades \cite{Khlopov:2008qy,Carr:2009jm,Carr:2020gox,Carr:2024nlv}. Not having formed from stellar collapse, they span a much wider mass range than their stellar cousins, and may have formed in the early universe from density perturbations. Their wide mass range makes them possible candidates of dark matter \cite{Belotsky:2014kca,Ballesteros:2017fsr,Clesse:2017bsw,Lehmann:2019zgt,Carr:2020xqk,Green:2020jor,Calza:2024fzo,Davies:2024ysj,Calza:2024xdh,Calza:2025mwn,Asmanoglu:2025agc,Dialektopoulos:2025mfz}, connecting them to new physics scenarios.

Historically, the formation of primordial black holes has been described by Carr density contrast criterion \cite{Carr:1975qj},
\begin{align}
\delta = \frac{\rho}{\rho_b} - 1 > \omega \, ,
\end{align}
where $\rho$ is the density of a perturbation, $\rho_b$ is the background density, and $\omega$ denotes the equation of state parameter. This criterion has since been refined, both analytically \cite{Kopp:2010sh,Harada:2013epa,Musco:2018rwt,Young:2019osy,Yoo:2022mzl} and numerically \cite{Niemeyer:1999ak,Shibata:1999zs,Hawke:2002rf,Musco:2004ak,Polnarev:2006aa,Musco:2008hv,Harada:2016mhb,Escriva:2019phb}. Loosely speaking, it corresponds to the cosmological extension of the hoop conjecture for black hole formation: if a mass $M$ is concentrated into a region that is smaller than its associated Schwarzschild radius $2GM/c^2$, collapse into a black hole is inevitable. A main output of such computations and estimates is the primordial black hole mass spectrum, which is highly relevant for realistic observational  constraints since it connects density perturbations (typically predicted from inflationary models, see e.g. Ref.~\cite{Ballesteros:2017fsr}) to the present-day abundance of primordial black holes.

If, however, primordial black holes are not properly described by the Schwarzschild metric and general relativity, an important question is: how does this affect the Carr criterion? In the absence of a concrete ultraviolet-complete gravitational theory, much activity has been devoted to the construction of so-called non-singular black hole models \cite{Bardeen:1968,Dymnikova:1992ux,Bonanno:2000ep,Hayward:2005gi,Frolov:2014jva,Frolov:2016pav,Carballo-Rubio:2025fnc}. In such settings, one introduces a regulator length scale $\ell > 0$ that parametrizes the length scale where deviations from general relativity are expected to occur. A central feature of non-singular models is the existence of a mass gap $M_\text{gap} \sim c^2\ell/G$, below which no black hole horizon exists \cite{Hayward:2005gi}. Dynamical calculations have confirmed this mass gap for black hole formation \cite{Frolov:2015bta}.

The goal of this paper is to extend these considerations to the cosmological setting, and explore how $\ell > 0$ affects the Carr density contrast criterion. This paper is organizes as follows: 

In Sec.~\ref{sec:non-singular-gravity} we begin by reviewing a simple model for non-singular gravity, drawing on recent work in the context of nonlocal field theory. Therein, we define effective densities that encode the presence of a gravitational regulator via associated smearing operations in non-singular gravity. We summarize known results for point masses, and introduce a spectral approach that allows us to derive effective densities of spherical shells and solid spheres. Then, in Sec.~\ref{sec:jeans-criterion}, utilizing the effective density of a solid sphere, we study its impact on the Jeans criterion, and find that the presence of a regulator increases the gravitational free-fall time, leading to a starkly different gravitational collapse scenario. Motivated by these results, in Sec.~\ref{sec:cosmology} we then construct a set of self-consistent Friedmann equations via a smeared out continuity equation, and use an effective potential approach to determine the dynamics of spherical shells in such a setting. We find, again, the existence of a mass gap $M_\text{gap}$, and relate it to the density contrast at horizon crossing, culminating in a modified Carr criterion for primordial black hole formation. We then summarize our approach, including its limitations, in Sec.~\ref{sec:conclusions}.

\section{Effective densities in non-singular gravity}
\label{sec:non-singular-gravity}

The singular Newtonian potential is a direct consequence of applying Gauss' law to a point-like particle,
\begin{align}
\nabla^2 \phi = 4\pi G M \delta{}^{(3)}(\ts{x}) \, .
\end{align}
This behavior can be ameliorated by introducing a higher-derivative modification according to the prescription
\begin{align}
\nabla^2 \rightarrow \nabla^2 f(\nabla^2) \, ,
\end{align}
where $f$ is usually referred to as a \emph{form factor} that affects the kinetic term of the theory. While such theories are motivated from fundamental insights in both string theory \cite{Frampton:1988kr} and asymptotic safety \cite{Knorr:2019atm} and have also been established as a separate approach \cite{Modesto:2011kw,Biswas:2011ar}, in this work we take a phenomenological approach and assume that the Newtonian limit of non-singular gravity can be parametrized in this way.

In order to make contact with principles from Newtonian gravity (and, later, cosmology) it is helpful to recast the modified Poisson equation
\begin{align}
\nabla^2 f(\nabla^2) \phi = 4\pi G \rho
\end{align}
into its standard form via an effective energy density,
\begin{align}
\nabla^2 \phi = 4\pi G \rho_\text{eff} \, , \quad \rho_\text{eff} = f^{-1}(\nabla^2) \rho \, ,
\label{eq:effective-density}
\end{align}
where we emphasize that in the present context of time-independent field configurations the operator $f^{-1}(\nabla^2)$ is well-defined since $f(\nabla^2) \not=0$ when acting on sources \cite{Boos:2020qgg}. In what follows, for concreteness we will focus on the nonlocal form factor\footnote{This choice is not unique. In fact, the conclusions of this paper would remain qualitatively unchanged if one were to choose a different form factor. However, the present form factor allows for many calculations to be performed analytically, which is why it is convenient for analytical studies. For more details on different form factors, we refer to Ref.~\cite{Boos:2020qgg} and references therein.}
\begin{align}
f(\nabla^2) = \exp\left( -\ell^2\nabla^2 \right) \, , \quad \ell > 0 \, ,
\end{align}
where $\ell > 0$ denotes the regulator scale, and in the limit of $\ell \rightarrow 0$ we recover the results of standard Newtonian theory. We define the spatial Green function via
\begin{align}
\nabla^2 f(\nabla^2) G(\ts{x}, \ts{x'}) = - \delta^{(3)}( \ts{x}-\ts{x'}) \, ,
\end{align}
and obtain for the above choice
\begin{align}
\label{eq:gf}
G(\ts{x},\ts{x'}) = -\frac{1}{4\pi |\ts{x} - \ts{x'}|} \text{erf}\left( \frac{|\ts{x}-\ts{x'}|}{2\ell} \right) \, ,
\end{align}
which is finite as $\ts{x} \rightarrow \ts{x'}$, encoding the UV improvement of such nonlocally modified theories provided $\ell \not= 0$. In what follows, we will briefly review the spectral decomposition of the radial Laplacian, and then determine explicit expressions for the gravitational potentials as well as the corresponding effective densities for the cases of a point mass, a spherical shell, and a solid sphere.

\subsection{Spectral decomposition}

The radial Laplacian, acting on a test function $\phi(r)$, takes the form
\begin{align}
\nabla_r^2 \phi(r) = \frac{1}{r^2} \frac{\partial}{\partial r} \left[ r^2 \frac{\partial}{\partial r} \phi(r) \right] \, .
\end{align}
Test functions, in this context, are functions that are (i) regular everywhere (in particular at $r=0$), and (ii) finite as $r \rightarrow \infty$. Eigenfunctions compatible with this ansatz are
\begin{align}
u_\lambda(r) = \sqrt{\frac{2}{\pi}} \frac{\sin\lambda r}{r} \, ,
\end{align}
such that
\begin{align}
\nabla_r^2 u_\lambda(r) = -\lambda^2 u_\lambda(r) \, .
\end{align}
These functions are normalized under the product
\begin{align}
\langle u_{\lambda}(r), u_{\lambda'}(r) \rangle \equiv \int\limits_0^\infty r^2 \dd r \, u_{\lambda}(r) u_{\lambda'}(r) = \delta(\lambda - \lambda') \, .
\end{align}
Test functions can then be expanded as
\begin{align}
\phi(r) = \int\limits_0^\infty \dd\lambda \, c_\lambda \, u_\lambda(r) \, , \quad c_\lambda = \langle \phi(r), u_\lambda(r) \rangle \, .
\end{align}
Two relevant examples include \cite{Hartenfels:2026}
\begin{align}
\label{eq:delta-rR}
\langle \delta(r-R), u_\lambda(r) \rangle &= \sqrt{\frac{2}{\pi}} R \sin \lambda R \, , \\
\label{eq:theta-rR}
\langle \theta(R-r), u_\lambda(r) \rangle &= \sqrt{\frac{2}{\pi}} \frac{\sin \lambda R - \lambda R \cos \lambda R}{\lambda^2} \, .
\end{align}

\subsection{Effective densities}
Let us now apply this method to derive the effective densities for a point mass, a hollow shell, and a solid sphere.

\subsubsection{Point mass}
For a point mass $M$ with density
\begin{align}
\rho = M \delta^{(3)}(\ts{x})
\end{align}
one obtains the effective density
\begin{align}
\rho_\text{eff} = \frac{M}{(4\pi\ell^2)^{3/2}} \exp \left[ - \frac{(\ts{x}-\ts{x}')^2}{4\ell^2}  \right] \, ,
\end{align}
corresponding to the well-known potential \cite{Modesto:2011kw,Biswas:2011ar,Edholm:2016hbt,Boos:2018bxf}
\begin{align}
\phi = -\frac{GM}{r} \text{erf} \left( \frac{r}{2\ell} \right) \, .
\end{align}
The effective density corresponds to a three-dimensional Gaussian profile of width $2\ell$, and its normalization ensures that in the limit of $\ell \rightarrow 0$ one recovers the three-dimensional delta function.

\subsubsection{Spherical shell}
Let us now consider a hollow sphere of mass $M$ and radius $R$, whose density is given by
\begin{align}
\label{eq:rho-hollow-sphere}
\rho = \frac{M}{4\pi R^2} \, \delta(r-R) \, .
\end{align}
Utilizing \eqref{eq:delta-rR} we find for the effective density
\begin{align}
\rho_\text{eff} = \frac{M}{8\pi^{3/2}\ell R r} \exp\left[-\frac{r^2+R^2}{4\ell^2} \right] \sinh \left( \frac{rR}{2\ell^2} \right) \, ,
\end{align}
giving rise to the potential
\begin{align}
\phi &= -\frac{GM}{r}\Bigg\{ \hspace{15pt} \frac12 \left(1 + \frac{r}{R} \right)\,\text{erf}\left(\frac{r+R}{2\ell}\right) \nonumber \\
& \hspace{54pt} + \frac12 \left(1 - \frac{r}{R} \right)\,\text{erf}\left(\frac{r-R}{2\ell}\right) \\
& - \frac{\ell}{\sqrt{\pi}R}\left[ \exp\left( -\frac{(r - R)^2}{4 \ell^2} \right) - \exp\left( -\frac{(r + R)^2}{4 \ell^2} \right) \right] \Bigg\} \, . \nonumber
\end{align}
This result matches the potential obtained from the hollow sphere mass density \eqref{eq:rho-hollow-sphere} and the Green function \eqref{eq:gf}, as expected.

\subsubsection{Solid sphere}
Finally, we focus on a solid sphere, whose density is\footnote{We note that since $\partial_x \theta(x-x_0) = \delta(x-x_0)$, the analytical problems of the spherical shell of radius $R$ and that of the solid sphere of the same radius are related. In fact, a similar analogy has been utilized by one of us in the context of relating the field of a point particle to that of an extended string \cite{Boos:2021suz}. However, in the present context this relation, while exact, does not lend itself to analytical simplifications, which is why we proceed with the spectral approach.}
\begin{align}
\rho = \frac{3M}{4\pi R^3}\, \theta(R-r) \, .
\end{align}
Using now Eq.~\eqref{eq:theta-rR}, the effective density can be calculated:
\begin{align}
\rho_\text{eff} &=  \frac{3M}{4\pi R^3} \bigg\{ 
\hspace{15pt} \frac12 \left[ \text{erf}\left( \frac{r+R}{2\ell} \right) - \text{erf}\left( \frac{r-R}{2\ell} \right) \right] \nonumber \\
&\hspace{50pt} + \frac{\ell}{\sqrt{\pi} r} \exp\left( -\frac{(r+R)^2}{4\ell^2} \right) \\
&\hspace{50pt} - \frac{\ell}{\sqrt{\pi} r} \exp\left( - \frac{(r-R)^2}{4\ell^2} \right) \bigg \} \, . \nonumber
\end{align}
The resulting potential takes the somewhat lengthy but exact form:
\begin{align}
\begin{split}
\phi &=-\frac{GM}{r} \Bigg \{ 
\hspace{11pt}        \frac12 \left[ \left( 1 - \frac{r}{R} \right)^2 \left( \frac{r}{2R} + 1 \right) \right] \text{erf}\left(\frac{r - R}{2 \ell}\right) \\
& \hspace{11pt} - \frac12 \left[ \left( 1 + \frac{r}{R} \right)^2 \left( \frac{r}{2R} - 1 \right)  \right] \text{erf}\left(\frac{r + R}{2 \ell}\right) \\
& \hspace{11pt} + \frac{\ell}{\sqrt{\pi}R}  \left[ \exp\left( -\frac{(r - R)^2}{4 \ell^2} \right) \left(1 + \frac{r}{2R} - \frac{r^2}{2R^2} \right) \right] \\
& \hspace{11pt} - \frac{\ell}{\sqrt{\pi}R} \left[ \exp \left( -\frac{(r + R)^2}{4 \ell^2} \right) \left(1 - \frac{r}{R} - \frac{r^2}{R^2}\right) \right] \\
&\hspace{11pt} + \frac{3\ell^2 r}{2R^3}  \left[ \text{erf}\left(\frac{r - R}{2 \ell}\right) -  \text{erf}\left(\frac{r + R}{2 \ell}\right) \right] \\
&\hspace{11pt} + \frac{2\ell^3}{\sqrt{\pi}R^3} \left[ \exp\left( -\frac{(r - R)^2}{4 \ell^2} \right) - \exp\left( -\frac{(r + R)^2}{4 \ell^2} \right) \right] \Bigg\} \, . \label{eq:phi-solid}
\end{split}
\end{align}
For later convenience we note that the average density of a solid sphere is
\begin{align}
\label{eq:rho-eff-sphere}
\bar{\rho}_\text{eff} &= \frac{3}{4\pi R^3} \int\limits_0^R \dd r \, 4\pi r^2 \rho_\text{eff} \\
&= \frac{3M}{4\pi R^3} \bigg\{ \text{erf}\left( \frac{R}{\ell} \right) - \frac{\ell}{\sqrt{\pi}R}\left[ 3 - \exp\left( -\frac{R^2}{\ell^2} \right) \right] \nonumber \\
&\hspace{83pt} + \frac{2\ell^3}{\sqrt{\pi}R^3} \left[ 1 - \exp\left( - \frac{R^2}{\ell^2} \right) \right] \bigg\} \, . \nonumber
\end{align}

\section{Modified Jeans criterion}
\label{sec:jeans-criterion}

\begin{figure}[!htb]
\centering
\includegraphics[width=0.5\textwidth]{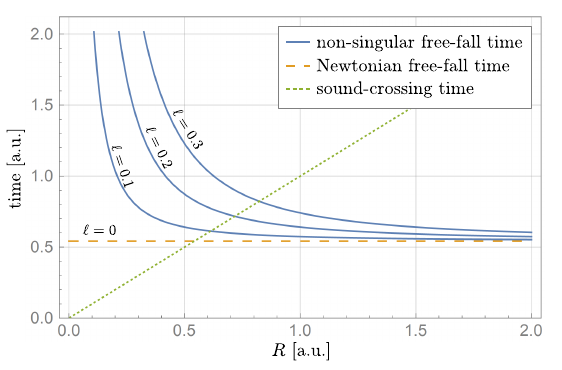} 
\caption{In this qualitative plot, we visualize the radius dependence of the free-fall time in non-singular gravity. For the sake of concreteness, we highlight three cases of varying regulator (solid lines) and compare their corresponding radius-dependent free-fall time to the Newtonian prediction (dashed line). The dotted line represents a typical sound-crossing time (here with speed of sound set to unity). The Jeans radii are determined as the intersection point of the sound-crossing time and the various free-fall times. For smaller regulator values, the resulting behavior at radii $R \gg \ell$ approaches the Newtonian result, whereas for smaller radii the free-fall time diverges, hinting towards a total avoidance of gravitational collapse, provided the regulator scale is much smaller than the radius of the considered matter distribution.}
\label{fig:jeans-criterion}
\end{figure}

In order to estimate the nature of gravitational collapse in he context of non-singular gravity, we will first study the Jeans criterion in this setting. To that end, we consider a solid sphere of radius $R$ of constant density $\rho$ with constant density, collapsing under its own gravity. This timescale is called the free-fall time, and is given by
\begin{align}
t_f(R) = \int\limits_0^R \frac{\dd r}{\sqrt{2[\phi_{R,\ell}(R) - \phi_{r,\ell}(r)}]} \, ,
\end{align}
where $\phi_{R,\ell}$ is the potential of a solid sphere of radius $R$, see Eq.~\eqref{eq:phi-solid}, and $\phi_{r,\ell}$ denotes Eq.~\eqref{eq:phi-solid} under the substitution $R \rightarrow r$,
\begin{align}
\phi_{r,\ell}(r) &= -\frac{GM}{r} \bigg\{ \left( 1 - \frac{3\ell^2}{2r^2} \right) \text{erf}\left( \frac{r}{\ell} \right) \\
& \hspace{52pt} + \frac{\ell}{\sqrt{\pi}r} \exp\left( -\frac{r^2}{\ell^2} \right) \\
& \hspace{52pt} + \frac{2\ell^3}{\sqrt{\pi}r^3}\left[ 1 - \exp\left( -\frac{r^2}{\ell^2} \right) \right] \bigg\}
\end{align}
In what follows, we parametrize $M = 4\pi R^3/3$ to describe a sphere of constant density, which is the usual setting in which the Jeans instability is discussed. Notice that since here we are working with the exact potential of a solid sphere \eqref{eq:phi-solid}, effective densities do not play a role. Last, we define the sound-crossing time as
\begin{align}
t_s = \frac{R}{c_s} \, ,
\end{align}
where $c_s$ denotes the speed of sound. Whenever the sound-crossing time exceeds the free-fall time, gravitational anisotropies can be averaged out by pressure waves. Conversely, gravitational collapse occurs if
\begin{align}
\label{eq:jeans-criterion}
t_f > t_s \, .
\end{align}
While the complicated analytical form of the potential \eqref{eq:phi-solid} prohibits an analytical study of the free-fall time, a numerical study of the above Jeans criterion for gravitational collapse, see Fig.~\ref{fig:jeans-criterion}, reveals that the presence of a regulator $\ell > 0$ leads to a radius-dependent free-fall time. Moreover, for $R/\ell \rightarrow 0$ the free-fall time diverges. This is physically expected, since gravitational interactions in non-singular gravity are regularized at length scales comparable to the regulator, indicating that gravitational collapse is halted at a certain scale.

This result is our key motivation to consider the above concepts embedded in a cosmological setting.

\section{Cosmological setting}
\label{sec:cosmology}

In our conventions, we describe the cosmological background as an FLRW metric via
\begin{align}
\dd s^2 = - \dd t^2 + a^2(t) \left[ \frac{\dd r^2}{1 - k r^2} + r^2 \dd\Omega^2 \right] \, ,
\end{align}
where in our conventions $r$ is dimensionless, such that $a(t)$ sets the physical scale, and $k=0,1$ describes a flat or closed universe, respectively. In order to describe gravitational collapse in such a setting, it is convenient to introduce the areal radius \cite{Harada:2013epa,Yoo:2022mzl}
\begin{align}
R(t) \equiv r a(t) \, .
\end{align}
Here, the dimensionless $r$ labels shells of different radii. Re-writing the Friedmann equations in this parametrization (here for vanishing cosmological constant, $\Lambda = 0$) one finds
\begin{align}
\label{eq:friedmann-gr}
\frac12 \dot{R}^2 = E - V_\text{eff} \, , \quad \ddot{R} = -\frac{4\pi G}{3}(\rho + P) R \, ,
\end{align}
with $\dot{} = \partial/\partial t$, and where $\rho$ and $P$ describe the density and pressure, respectively, related by the continuity equation
\begin{align}
\dot{\rho} = -3 H \left( \rho + P \right) \, , \quad H = \frac{\dot{R}}{R} \, .
\end{align}
We moreover defined
\begin{align}
V_\text{eff} = -\frac{4\pi G}{3} \rho R^2 \, , \quad E = -\frac{kr^2}{2} \, .
\end{align}
This allows the qualitative description of shell dynamics via the condition\footnote{We note that the occurrence of $r$ in $E = -kr^2/2$ is merely a gauge choice and in what follows we will hence set $r=1$ such that $E = -k/2$.}
\begin{align}
V_\text{eff} \leq E \, .
\end{align}
In the Newtonian case, $\rho = 3M/(4\pi R^3)$, and one recovers the usual Coulomb-style dynamics via $V_\text{eff} = -GM/R$. In this case, as is well known, gravitational collapse can never be prevented. However, if the effective potential were to exhibit a minimum, dependent on a combination of $M$ and the regulator $\ell$, a scenario where no collapse can occur becomes feasible.

\subsubsection{Modified Friedmann equations}

However, in the absence of a fundamental theory, it is not clear what constitutes the equivalent of the Friedmann equations in non-singular gravity. As a heuristic approach, we notice that the effect of the regulator is a ``smearing out'' of the gravitational field. This idea has been discussed in nonlocal gravity \cite{Boos:2018bhd,Giacchini:2018wlf} as well as more recently in relation to topological invariants \cite{Boos:2024sgm}: Namely, it has been shown that Gauss-type flux integrals of monopole charges no longer correspond to conserved quantities, leading to an altered continuity equation of the form
\begin{align}
\label{eq:cont-modified}
\dot\rho + 3 H (\rho + P) = \Gamma \, ,
\end{align}
where $\Gamma \not= 0$ is a ``leakage term'' that parametrizes the smearing of energy distributions at a characteristic length scale of the regulator $\ell > 0$. As $\ell \rightarrow 0$, one has $\Gamma \rightarrow 0$ and recovers the standard continuity equation. Now, we take this leakage term to indicate that the first Friedmann equation can be written in a similar form to the general relativistic one, under the change $E \rightarrow \tilde{E}(R)$, where $\tilde{E}(R)$ is now an $R$-dependent ``energy'' of a shell of size $R$. We emphasize that this, unlike the appearance of $r$ in $E$ before, is \emph{not} a gauge choice, but rather a physical consequence embodying the smearing out of energy on a characteristic regulator length scale $\ell > 0$. Again, in the limit of $\ell \rightarrow 0$ we expect that $\tilde{E}(R) \rightarrow E = -k/2$. Last, this set of assumptions will be sufficient to recover the second Friedmann equation, which is why refer to this approach as ``self-consistent.'' Let us now prove this.

We begin by stipulating the form of the first Friedmann equation,
\begin{align}
\frac12 \dot{R}^2 = \tilde{E}(R) - V_\text{eff}(R) \, ,
\end{align}
and then differentiate with respect to cosmic time $t$ and utilize the continuity equation \eqref{eq:cont-modified} such that
\begin{align}
\ddot{R} = - \frac{4\pi G}{3}(\rho + 3P) R + \tilde{E}'(R) + \frac{4\pi G}{3} \frac{R^2}{\dot{R}}\Gamma \, ,
\end{align}
where $\tilde{E}'(R) = \partial_R \tilde{E}(R)$. This corresponds to the second Friedmann equation, provided
\begin{align}
\tilde{E}'(R) + \frac{4\pi G}{3}\frac{R^2}{\dot{R}} \Gamma = 0 \, .
\end{align}
This, using the continuity equation  and assuming now that all time dependence of the density function enters via $R(t)$ [so we can write $\dot{\rho} = (\partial_R \rho) \dot{R}$] gives the following differential equation for $\tilde{E}(R)$:
\begin{align}
\tilde{E}'(R) = -\frac{4\pi G}{3} \left[ \rho' R^2 + 3R(\rho+P) \right] \, .
\end{align}
This ordinary differential equation can be directly integrated, yielding
\begin{align}
\label{eq:etilde}
\tilde{E}(R)-\tilde{E}(R_\text{max}) = \frac{4\pi G}{3} \int\limits_{R_\text{max}}^0 \dd R \left[ \rho' R^2 + 3R(\rho+P) \right] \, .
\end{align}
This then implies the generalized Friedmann equations,
\begin{align}
\begin{split}
\frac12 \dot{R}^2 &= \tilde{E}(R) - V_\text{eff}(R) \, , \\
\ddot{R} &= - \frac{4\pi G}{3}(\rho + 3P) R \, .
\end{split}
\end{align}
In what follows, we will assume an equation of state,
\begin{align}
P = \omega \rho \, ,
\end{align}
and in order to establish some physical intuition behind these equations we will focus on the average effective density of a solid sphere of radius $R$ and mass $M$ in presence of a regulator $\ell > 0$, which we briefly state here again takes the form
\begin{align}
\bar{\rho}_\text{eff} &= \frac{3M}{4\pi R^3} \bigg\{ \text{erf}\left( \frac{R}{\ell} \right) - \frac{\ell}{\sqrt{\pi}R}\left[ 3 - \exp\left( -\frac{R^2}{\ell^2} \right) \right] \nonumber \\
&\hspace{55pt} + \frac{2\ell^3}{\sqrt{\pi}R^3} \left[ 1 - \exp\left( - \frac{R^2}{\ell^2} \right) \right] \bigg\} \, . \tag{\ref*{eq:rho-eff-sphere}}
\end{align}
Notice that in the limiting case $R \gg \ell$ we recover the constant density of a sphere of radius $R$ and mass $M$.

\subsubsection{Infinite universe}

Embedded in an infinite universe we set $R_\text{max} = \infty$ as well as $E(R_\text{max}) = -1/2$ in Eq.~\eqref{eq:etilde} and obtain
\begin{align}
\tilde{E}(R) = -\frac12 + \frac{3GM\ell}{2\sqrt{\pi}R^2}\left\{ 1 + \left[ 1 - \exp\left( -\frac{R^2}{\ell^2} \right) \frac{\ell^2}{R^2} \right] \right\} \, ,
\end{align}
For small distances $R \ll \ell$ we find
\begin{align}
\tilde{E} - V_\text{eff} = -\frac12 + \frac{3GM}{4\sqrt{4}\ell} + \mathcal{O}(R^2) \, .
\end{align}
Collapse becomes impossible if $V_\text{eff} > E(R)$ for all values of $R$, which in this case corresponds to
\begin{align}
GM < GM_\text{gap} = \frac{2\sqrt{\pi}\ell}{3} \approx 1.18 \ell \, .
\end{align}
The existence of this mass gap $M_\text{gap}$ demonstrates that in the presence of a regulator $\ell > 0$ gravitational attraction becomes weaker at short scales, and therefore a critical mass $M_\text{gap}$ is required to induce gravitational collapse, in stark contrast to Newtonian gravity.

\subsubsection{Carr criterion}

We are now ready to embed this collapse criterion in a simplified model of the early Universe to study its implications for the formation criterion on primordial black holes via density fluctuations. We assume the background universe evolution to be flat ($k=0$) such that the background density $\rho_b$ is given directly from the unmodified Friedmann equation via
\begin{align}
\rho_b &= \frac{3 H^2}{8\pi G} \, .
\end{align}
We now assume the existence of a density perturbation
\begin{align}
\delta = \frac{\rho}{\rho_b} - 1 \, ,
\end{align}
stemming from an external mechanism (e.g. primordial density fluctuations) during radiation domination, occurring roughly during
\begin{align}
10^{-32} \, \text{s} \lesssim t \lesssim 10^{12} \, \text{s} = 50,000 \, \text{years} \, ,
\end{align}
At that cosmological epoch, the horizon is located at
\begin{align}
\label{eq:rh}
R_H = \frac{1}{H} = 2 c t \, .
\end{align}
Physically, this implies that perturbations whose wavelengths are smaller than $R_H$ can re-enter the horizon, after it has been shrunk during inflation, and then potentially collapse and form black holes. In this context, it is useful to introduce the concept of a horizon mass,
\begin{align}
M_H = \frac{4\pi}{3} \rho_H R_H^3 \, .
\end{align}
During pure radiation domination one finds
\begin{align}
M_\text{H}(t) = \frac{4\pi}{3} R_\text{H}^3 \rho(t) = \frac{c^3 t}{G} \, ,
\end{align}
implying a total mass range for primordial black holes of the order
\begin{align}
4 \times 10^{-8} \, \text{kg} < M_\text{H} < 4 \times 10^{36} \, \text{kg} \, .
\end{align}
However, the existence of perturbations can significantly change these estimates. At horizon entry, $\rho = \rho_H$ and $H = 1/R_H$ such that
\begin{align}
\label{eq:deltaH}
\delta_H \equiv \frac{\rho_H}{\rho_b} - 1 = \frac{3M_H}{4\pi R_H^3} \frac{1}{\frac{3}{8\pi G} H^2} - 1 = \frac{2GM_H}{R_H} - 1
\end{align}
Bearing in mind the previous discussion from the infinite-universe case, we expect that the existence of a mass gap $M_\text{gap}$ for gravitational collapse then immediately generates a minimum threshold for $\delta_H$, quite similar to the Carr criterion.

\begin{figure*}[!htb]
\centering
\includegraphics[width=0.48\textwidth]{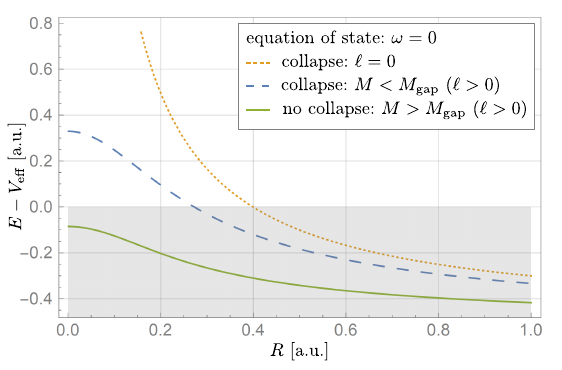} 
\includegraphics[width=0.48\textwidth]{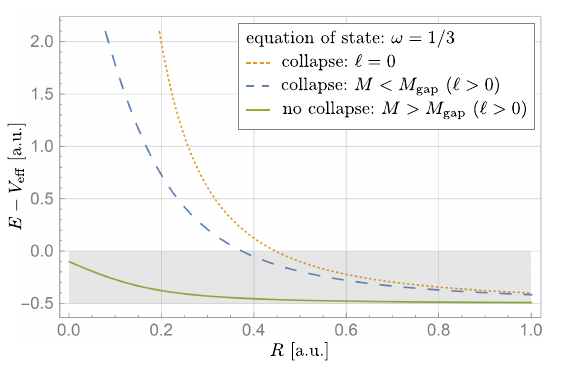} 
\caption{The kinematic condition $V_\text{eff}(R) \leq E$, derived from the first Friedmann equation \eqref{eq:friedmann-gr}, is analyzed graphically for dust (left panel) and radiation (right panel). The shaded region corresponds to the kinematically forbidden region $V_\text{eff} > E$. The Newtonian case corresponds to the dotted line, and the non-singular cases are described by the dashed and solid lines. The behavior in both cases is identical: In the case of vanishing regulator, collapse will always occur for small enough radii. In the presence of the regulator, this behavior is shifted towards qualitatively smaller radii. If a critical mass threshold is exceeded, however, collapse will always be avoided.}
\label{fig:collapse}
\end{figure*}

\begin{figure}[!htb]
\centering
\includegraphics[width=0.5\textwidth]{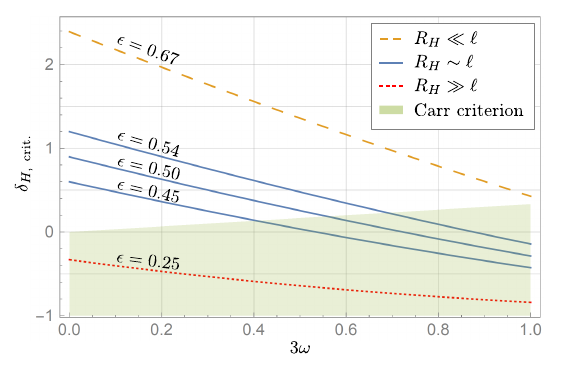} 
\caption{The Carr criterion for the density contrast required for black hole formation (shaded area excluded) is compared to the mass gap-induced collapse criterion. If the horizon scale $R_\text{H}$ is larger or of the same order of the regulator size $\ell$, the new constraints are tighter than the Carr criterion; if they are of comparable sizes, then the new constraints affect smaller $\omega$ values. As expected, for $\ell \ll R_\text{H}$ the Carr criterion is more stringent than the mass-gap induced collapse criterion presented in this paper.}
\label{fig:carr}
\end{figure}

To determine these implications, let us focus again on the effective density of a solid sphere, see Eq.~\eqref{eq:rho-eff-sphere}, but we will allow for an equation of state and therefore multiply this density with the cosmological redshift factor $(R_H/R)^{3\omega}$, arriving at
\begin{align}
\rho &= \frac{3M}{4\pi R^3} \bigg\{ \text{erf}\left( \frac{R}{\ell} \right) - \frac{\ell}{\sqrt{\pi}R}\left[ 3 - \exp\left( -\frac{R^2}{\ell^2} \right) \right] \\
&\hspace{25pt} + \frac{2\ell^3}{\sqrt{\pi}R^3} \left[ 1 - \exp\left( - \frac{R^2}{\ell^2} \right) \right] \bigg\} \, \times \, \left(\frac{R_H}{R}\right)^{3\omega} \, . \nonumber
\end{align}
Unlike before, we now embed this density inside a finite, closed universe, and normalize the energy such that $E(R_H) = -1/2$. The effective potential is defined as before,
\begin{align}
V_\text{eff} = - \frac{4\pi G}{3} \rho R^2 \, .
\end{align}
Then, Eq.~\eqref{eq:etilde} can again be integrated, and we obtain the somewhat lengthy expression
\begin{align}
\begin{split}
\tilde{E}(R) &= \hspace{11pt} \frac{3GM}{4\sqrt{\pi}R} \left(\frac{R_H}{R}\right)^{3\omega} h\left( \frac{R}{\ell} \right) \\
&\hspace{8pt}- \frac{3GM}{4\sqrt{\pi}R_H} h\left( \frac{R_H}{\ell} \right) - \frac12 \, , \\
h(q) &= \frac{1}{q} \bigg[ \hspace{11pt} \frac{4}{2+3\omega} + \frac{2}{q^2} \left( e^{-q^2} - \frac{4}{4+3\omega} \right) \\
&\hspace{25pt}- \frac{3\omega}{q^2} \text{Ei}\left( 3 + \frac{3\omega}{2}, q^2 \right) \bigg] \, ,
\end{split}
\end{align}
where $\text{Ei}(n,z)$ denotes the exponential integral
\begin{align}
\text{Ei}(n,z) \equiv \int\limits_1^\infty \dd t \frac{e^{-zt}}{t^n} \, .
\end{align}
Analyzing the behavior of $\tilde{E}(R)-V_\text{eff}(R)$ at small shell radii $R \ll \ell$, we find that this quantity is not always positive, implying that for masses below a critical threshold, no collapse occurs. This exactly reproduces the behavior found in the infinite-universe context from above, and we visualize this in Fig.~\ref{fig:collapse} for the case of dust ($\omega=0$) and radiation ($\omega=1/3$). While in general it is possible to extract the $\omega$-dependent mass gap values from the formulas above,
\begin{align}
\label{eq:mgap}
GM_\text{gap}^\omega &= \frac{2\sqrt{\pi}\ell}{3} \bigg\{ -\frac{4\epsilon^2}{2+3\omega} + \frac{8\epsilon^4}{4+3\omega} - 2\epsilon^4 e^{-1/\epsilon^2} \\
& -3 \omega \epsilon^{-3\omega} \left[ \Gamma\left( -2 -\tfrac{3\omega}{2} \right) - \Gamma\left( -2-\tfrac{3\omega}{2}, \tfrac{1}{\epsilon^2} \right)\right] \bigg\}^{-1} \, , \nonumber \\
\epsilon &= \frac{\ell}{R_H} \, , 
\end{align}
here we provide them for the dust case and the radiation case as well explicitly,
\begin{align}
GM_\text{gap}^{0} &= \frac{2\sqrt{\pi}\ell}{3\left( 1 - 2\epsilon^2 + 2\epsilon^4 \right) - 6 e^{-1/\epsilon^2} } \, , \\
GM_\text{gap}^{1/3} &= \frac{10\sqrt{\pi}\epsilon \ell}{8\sqrt{\pi} - 20\epsilon^3 + \epsilon^5 \left[24 + 15 \text{Ei}\left(\tfrac72,\tfrac{1}{\epsilon^2}\right) - 30 e^{-1/\epsilon^2} \right] } \, . \nonumber
\end{align}
Crucially, via Eq.~\eqref{eq:deltaH}, this mass gap directly translates in to a critical threshold for the density contrast,
\begin{align}
\hspace{-8pt}
M_H > M_\text{gap} \quad \Leftrightarrow \quad \delta_H > \delta_{H,~\text{crit.}} \equiv \frac{2G M_\text{gap}}{R_H} - 1 \, .
\end{align}
Using the dimensionless parameter $\epsilon$ we then plot this new criterion as a function of the equation of state parameter $\omega$, and compare it to the Carr criterion, see Fig.~\ref{fig:carr}.

By considering $R_H \sim \ell $, equation \eqref{eq:rh} then allows us to estimate which cosmological epochs are sensitive to a given regulator scale,
\begin{align}
t \sim \frac{\ell}{2c} \, ,
\end{align}
For example, a femtometer regulator corresponds to
\begin{align}
t \sim 10^{-24}\,\text{s} \sim 10^6 \,\text{TeV} \, ,
\end{align}
but in the absence of a concrete model that parametrizes or predicts the density contrast, we cannot make more refined predictions. However, given a fundamental regulator scale $\ell$ and a primordial black hole mass $M$, we can infer at what horizon radius (and, therefore, at which time) this object must have formed.

\section{Conclusions}
\label{sec:conclusions}

We have presented a heuristic analysis of gravitational collapse criteria in the presence of a gravitational regulator with a focus on primordial black hole formation. In a first step, we reported on the existence of a mass gap $M_\text{gap}$, below which no gravitational collapse occurs, via two methods: a kinematic analysis of an effective Friedmann equation, as well as a numerical study of free-fall times in connection with the Jeans criterion for gravitational collapse. Second, we extended our considerations to the cosmological setting by computing the mass gap in a closed-universe approach. As a final step, we related this to a critical threshold in the local density contrast at the time of horizon crossing. This formally takes the same form as Carr's well-known criterion on primordial black hole formation, but depending on the size of the regulator $\ell$ in comparison to the horizon scale $R_\text{H}$ (the first of which is a free parameter, in principle, whereas the latter one is essentially the temperature of the universe), this new criterion may take precedence over the Carr criterion over a wide range of equation of state parameters. Viewed from the perspective of the mass gap, these considerations strongly imply that in the presence of ultraviolet gravitational regulators, gravitational collapse (and, therefore, primordial black holes themselves) are expected to form at threshold, namely
\begin{align}
M_\text{PBH} \gtrsim M_\text{gap}(\ell, R_H) \, .
\end{align}
Since the mass gap, see Eq.~\eqref{eq:mgap}, is monotonically decreasing in $R_H$ (keeping $\ell$ is fixed), the largest possible black hole mass is dictated by the smallest horizon scale $R_H$, establishing an intricate link of primordial black hole masses one one side, and new physics (such as UV-complete gravitational theories) on the other.

Clearly, the approach presented in this paper has its limitations. First, no fundamental field equations have been utilized. Rather, based on the Friedmann equations of general relativity, we developed a self-consistent approach wherein the two Friedmann equations follow from each other via a suitable generalized continuity equation that utilizes the interpretation of regulators as smearing operators. A fundamental theory of gravity could address this problem more rigorously.

Second, while the connection between early universe physics and primordial black hole formation is plausible, here we merely established a modification of the Carr criterion. If one takes an inflationary model and were to compute the density perturbations, then any observed primordial black hole mass would constrain the regulator $\ell$, and vice versa. A crucial step into this direction is the determination of a realistic mass spectrum for the black hole formation mechanism in this paper.

Third, the Carr criterion is not accurate---rather, it served as a historical estimate, that has since been improved both analytically and numerically. We introduce the mass gap and corresponding modified Carr criterion in the same spirit. We do not expect this estimate to be any more accurate than $\mathcal{O}(1)$, but nevertheless emphasize that in this paper we are trying to make a qualitative point, rather than a quantitative one: if a gravitational regulator exists, there should be a minimal mass, beyond no primordial black holes have formed.

\section{Acknowledgements}

JB is grateful for support as a Fellow of the Young Investigator Group Preparation Program, funded jointly via the University of Excellence strategic fund at the Karlsruhe Institute of Technology (administered by the federal government of Germany) and the Ministry of Science, Research and Arts of Baden-W\"urttemberg (Germany).

\end{document}